   \documentstyle[12pt,epsf]{article}
\begin{document}
  \baselineskip 15.2 pt \parskip 4 pt
  \begin{center}
   \vspace{-1.1 cm}
        {\it 1st International Workshop on Semiconducting and
	 Superconducting Materials\/}\\ Turin, 17-19 February 1999,
	 to be published in {\it Phil.\ Mag.\ B} \\[.8 cm]
  {\large\bf AC Response of Thin Film Superconductors at\\[.1cm]
             Various Temperatures and Magnetic Fields}
       \\[0.7 cm]
  Ernst Helmut Brandt \\[0.1 cm]
  Max-Planck Institut f\"ur Metallforschung, D-70506 Stuttgart, Germany
  \end{center}   \vspace{0.6 cm}
                 \noindent

  {\it Abstract:\/} A common method for measuring the electromagnetic
properties of superconductors is to measure their complex magnetic
ac susceptibility $\chi$ as a function of frequency $\omega$ and
amplitude $H_0$, and of temperature $T$ and applied dc magnetic
field $H_a^{dc}$ as parameters.
The basic theory of the linear $\chi(\omega)$ and nonlinear
$\chi(H_0, \omega)$ is outlined for various geometries, e.g.\ disks,
rings, and strips of thin films or thicker platelets in a perpendicular
magnetic field. It is shown how $\chi(\omega)$ explicitly
depends on the linear resistivity $\rho_{ac} (\omega) = E/J$ or
penetration depth $\lambda_{ac}(\omega)$, and $\chi(H_0, \omega)$
on the nonlinear current--voltage law $E(J,B)$ where $E({\bf r})$,
$J({\bf r})$, $B({\bf r})$ are the local electric field, current
density, and induction. The dependence of $E(J,B)$ on $T$ and on
various material properties like pinning forces or pinning
energies, structural defects and granularity,  leads to an
implicit dependence of $\chi$ on these parameters.
  \\[.5cm]
 {\bf \large 1. Introduction}

   The magnetic response of type-II superconductors may be considered
under various aspects. {\it First\/}, one may be interested in the
{\it microscopic} motion of the Abrikosov vortices, i.e.\ in
their pinning, thermally activated depinning, crossing of surface
barriers, and mutual annihilation.  {\it Second\/}, within
{\it continuum} theory these microscopic processes
are described by a current-voltage characteristic  ${\bf E=E(J,B)}$
(the local electric field caused by the current density ${\bf J}$
and depending also on the local induction ${\bf B}$) and by a
reversible magnetization curve ${\bf H=H(B)}$, which in sufficiently
large $B$ may be approximated by ${\bf H = B}/\mu_0$, but at smaller
$B$ may lead to a geometric barrier for flux penetration
(Zeldov {\it et al.} 1994, Brandt 1999).
In general, these material laws are anisotropic when the material is
anisotropic, when anisotropic defects are introduced, or when the Hall
effect is taken into account. In the isotropic case one has
${\bf H}=H(B) {\bf H}/H$ and (in simple geometries)
${\bf E}= \rho(J,B) {\bf J}$ with $\rho =E/J$.
{\it Third\/}, the {\it geometry} of the experiment, i.e.\ the
specimen shape and orientation of the applied magnetic field and/or
transport current have to be accounted for. The geometry determines
the local profiles ${\bf B(r)}$ and ${\bf J(r)}$ and the global
response, i.e.\ the magnetic moment and voltage drop. In some
cases the geometry modifies the appropriate law ${\bf E(J,B)}$.
For example, in isotropic films the critical current density
$J_c$ becomes anisotropic when a strong magnetic field is applied
parallel to the film
(Schuster {\it et al.} 1997).
{\it Fourth\/}, one may consider the vortex state as a ``phase'' which
undergoes transitions from a strong pinning state (``vortex glass'')
to a thermally depinned state when in the $BT$ plane an
``irreversibility line'' $T_{rev}(B)$ or ``glass line'' $T_g(B)$ is
crossed (Blatter {\it et al.} 1993, Brandt 1995),
or which changes from solid to ``liquid'' at a
``melting line'' $T_m(B)$ where the flux-line lattice melts
(Zeldov {\it et al.} 1995,  Majer {\it et al.} 1995,
 Schilling {\it et al.} 1996,  Welp {\it et al.} 1996,
 Roulin {\it et al.} 1998, Sasagawa {\it et al.} 1998,
 Dodgson {\it et al.} 1998).
In such a ``universal'' description
the detailed microscopic behavior of the vortices remains less clear;
in particular, the microscopic interpretation of the length and time
scales which diverge at the glass temperature $T_g(B)$ is still open.

  At low temperatures $T$, the current--voltage curve $E(J,B)$ is
highly nonlinear, typically $E=E_c\cdot(J/J_c)^n$ with large exponent
$n(B,T) \gg 1$, and one arrives at the usual Bean model
(Bean 1964, Campbell and Evetts 1972)
with critical current density $J_c(B,T)$. With increasing $T$, the
exponent $n$ decreases and one observes flux creep, i.e.\
nonlinear relaxation of currents and magnetic moment, see Sec.\ 2.
At high $T$ above the irreversibility line, thermal depinning leads to
a linear $E(J)$ which in general is frequency dependent and complex.
An interesting task is the detailed measurement of $E(J,B)$ of
thin superconductor films at various temperatures. A nontrivial
reversible $H(B)$ with a jump at the lower critical field
$H_{c1}$ (calculated, e.g.\ from the
Ginzburg-Landau equations) leads to surface barriers, which may be
microsopic as predicted by Bean and Livingston (1964), or geometric,
caused by the non-ellipsoidal shape of the superconductor.

 Some geometry problems have been solved exactly, e.g.\ the Bean model
for long superconductors in parallel field (Bean 1964) and for thin
disks (Mikheenko and Kuzovlev 1993) and long strips
(Brandt {\it et al.} 1993) in perpendicular field, and in strips with
transport current without (Norris 1970) and with
(Brandt and Indenbom 1993, McElfresh {\it et al.} 1994)
applied magnetic field. Known are further the linear
(Clem {\it et al.} 1976, Kes {\it et al.} 1989)
and nonlinear            (Rhyner 1993, Gilchrist and Dombre 1994)
relaxation (or flux diffusion) in longitudinal
geometry where demagnetization effects are absent, and the linear
and nonlinear reversible response of homogeneous ellipsoids, which
is obtained from the longitudinal results by introducing a
demagnetization factor. The magnetic response of thin strips and
disks (Brandt 1994a,b) in a perpendicular field is
easily computed by solving a one-dimensional (1D)  integral equation
for the current density, while thin rectangular (or arbitrarily shaped)
films and thick strips and disks  are two-dimensional (2D) problems,
which still can be computed on a PC (Brandt 1995a, 1996b, 1998).

  Recently two different algorithms were presented
(Labusch and Doyle 1997, Doyle {\it et al.} 1997, Brandt 1999)
which account for both constitutive laws ${\bf E=E(J,B,r)}$ and
${\bf H=H(B,r)}$ and which apply to all geometries, i.e.\ to
superconductors of any shape. These universal
methods allow to compute the geometric ``edge barrier'',
(Zeldov {\it et al.} 1994), e.g.\ in
strips and circular disks (or cylinders) of finite
thickness in perpendicular field. For the reversible magnetization
curve $H(B)$  any model may be used which approximates the exact
Ginzburg-Landau result (Brandt 1997a). If one is interested in the
behavior not too close to the upper critical field  $H_{c2}$, one
may use a simple hyperbola:
 $H(B) \approx [H_{c1}^2 +(B/\mu_0)^2]^{1/2} {\rm sign}(B)$,
for $B > \mu_0 H_{c1}$ and with undefined reversible field
 $-H_{c1} \le H \le H_{c1}$ at $B=0$.

  As mentioned above, really 3D problems (e.g.\ the
superconducting cube) have not been computed so far.
The difficulty is not only of
numerical nature, but it is even not clear which material laws
${\bf E(J,B)}$ are reasonable when ${\bf J}$ is not parallel to
${\bf B}$. One may introduce two critical current densities
$J_{c \perp}$ and $J_{c \|}$ for currents $\perp B$ and $\| B$.
In general, at low $B \ll B_{c2}$ (the upper critical field) one
should have $E\propto B$ since each vortex contributes independently
to the resistivity.

  Strictly spoken, even the flux penetration into thin films is a
3D problem if the specimen is not of exactly circular shape or is
not an infinitely long strip (Mikitik and Brandt 1999). This is so
since during the penetration of flux, the orientation of the sheet
current ${\bf J}_s(x,y) = \int {\bf J}(x,y,z) \,dz$ changes with
time. Therefore, the orientation of the vortices which penetrate from
the flat surfaces also changes, since these vortices are perpendicular
to ${\bf J}_s$. This means that at any point $x,y$ away from the
symmetry axes and from the specimen edge or center, in principle
the vortex orientation may be a function of the depth $z$,
i.e.\ the vortex arrangement may be twisted. This rotation is
most pronounced at moderate magnetic fields; it is absent when the
applied field is above the field of full penetration,
or when the film thickness $d$ is smaller than the magnetic
penetration depth $\lambda$, since in these cases the vortices
are nearly along $z$.
  \\[.5cm]
 {\bf \large 2. Definition of AC Susceptibilities}

  The global magnetic response of a superconductor of arbitrary
shape in a time dependent homogeneous magnetic field $H_a(t)$ may be
characterized by its (dipolar) magnetic moment
\begin{eqnarray}  
  {\bf m}(t) = {1 \over 2} \int {\bf r \times J(r},t) \, d^3r \,,
\end{eqnarray}
were ${\bf J(r},t)$ is the current density inside the specimen of
volume $V$. Extensions to inhomogeneous ${\bf H_a(r},t)$ and to
higher (e.g.\ quadrupolar) moments usually are not required when
an experiment is properly designed to obtain the material laws of
the specimen. These material laws may be linear or nonlinear. In
the {\it linear} case, the conductor is completely described by its
linear resistivity  $\rho_{ac}(\omega) =i\omega\mu_0\lambda^2_{ac}$,
which in general is complex and frequency dependent, or equivalently,
by a conductivity $\sigma_{ac}(\omega)=1/\rho_{ac}(\omega)$
or a complex penetration depth
$\lambda_{ac}(\omega) =(\rho_{ac}/i\omega\mu_0)^{1/2}$.
For example, an applied field $H_a(t)=H_0 e^{i\omega t}$
generates a local current density
${\bf J(r},t)={\bf J_0(r}) e^{i\omega t}$ and an electric field
${\bf E(r},t)={\bf E_0(r}) e^{i\omega t}$ where ${\bf J_0(r)}$
and ${\bf E_0(r)} = \rho_{ac} {\bf J_0(r)}$ are complex
amplitudes. The magnetic field then penetrates into a conducting
half space $x>0$ as $H(x,t)=H_a(t)\exp[-x/\lambda_{ac}(\omega)]$.
For magnetic materials with linear permeability
$\mu(\omega)=B/(\mu_0 H)$, in these formulae $\mu_0$ has to be
replaced by $\mu(\omega)\mu_0$. A superconductor may exhibit such
a linear response above a temperature where thermal depinning of
vortices, or thermally assisted flux flow (TAFF) occurs
(Kes {\it et al.} 1989).

   If pinning is strong, or if the reversible magnetization curve
$H=H(B)$ with a finite lower critical field $H_{c1}$ is accounted
for, the magnetic response is {\it nonlinear}. At sufficiently
large induction $B \gg \mu_0 H_{c1}$ one may put
${\bf B} = \mu_0 {\bf H}$. In this case pinning and flux creep may
be modeled  by a nonlinear resistivity law ${\bf E}=\rho(J){\bf J}$,
e.g.\ by the model $\rho(J) = \rho_0 (J/J_c)^{n-1}$ with critical
current density $J_c(B)$ and creep exponent $n(B) >1$, as mentioned
above. This model results when in the electric field
$E(J) = E_0 \exp[-U(J)/kT]$ an activation energy of the form
$U(J) = U_0 \ln(J_c/J)$ is inserted, yielding $E=E_c (J/J_c)^n$
with $n=U_0/kT$.

  Interestingly, when $E\propto J^n$ then during creep in the fully
penetrated state, after
some transient time $t_1$ the electric field exhibits universal
behavior, i.e.\ the ${\bf r}$ and $t$ dependences separate:
$E({\bf r},t) \propto f({\bf r})/(t_1+t)^{n/(n-1)} \approx
 f({\bf r})/t$, where the profile $f({\bf r})$ depends only on
the specimen shape (Gurevich and Brandt 1994, Brandt 1996a).
As a consequence,  $J({\bf r},t) \propto  f({\bf r})^{1/n}/
 (t_1+t)^{1/(n-1)} \propto {\rm const} -\ln t$ if $n \gg 1$,
and thus the magnetic moment relaxes as
$m \propto {\rm const} -\ln t$.
During flux creep the magnetic response to a small ac magnetic
field is {\rm linear}, though the underlying constitutive law
 $E \propto J^n$ is highly nonlinear (Gurevich and Brandt 1997,
Brandt and Gurevich 1996).

   From the calculated current density $J({\bf r},H_a)$ the
magnetic moment
${\bf m}(H_a)$ (1) is obtained. If $H_a(t)$ is cycled sinusoidally,
$H_a(t)=H_0 \sin(\omega t)$, then $m(t)$ (the component of ${\bf m}$
along $H_a$) performs a hysteresis loop from which {\it nonlinear}
complex ac susceptibilities
$\chi_\mu (H_0, \omega) =\chi'_\mu -i\chi''_\mu$
($\mu=1,$ 2, $3, \dots$) may be defined as
  \begin{eqnarray} 
 \chi_\mu (H_0,\omega) ={i \over \pi H_0} \int_0^{2\pi}\!\! m(t)\,
   e^{-i \mu \omega t} \, d(\omega t) \,.
  \end{eqnarray}
For small amplitudes $H_0 \to 0$ this yields $\chi_1(H_0,\omega)=
 m'(0) = \lim_{H_a \to 0} \partial m(H_a)/\partial H_a$.
A convenient normalization of the $\chi_\mu (H_0,\omega)$ is thus
$\chi_\mu \to \chi_\mu /|m'(0)|$, which at $H_0 \to 0$ yields
the ideal diamagnetic susceptibility $\chi_1(0,\omega) = -1$
if $m'(0) <0$. In the critical state model, with general
$J_c(B)$ or with Bean's $J_c=$const, the susceptibilities
$\chi_\mu(H_0,\omega)$ do not depend on $\omega$, i.e.\ all
cycling rates give the same loop $m(H_a)$ and same $\chi_\mu(H_0)$.

  In the opposite limit of {\it linear} response, no higher
harmonics are generated, i.e.\  $\chi_\mu =0$ for $\mu \ne 1$,
and $\chi_1 = \chi(\omega)$ does not depend on the amplitude. In
complex notation with $H_a(t)=H_0 e^{i\omega t}$  (only the real
parts of $H_a$, $H$, $B$, $E$, $J$, and $m$ have physical meaning)
the {\rm linear} ac susceptibility $\chi(\omega)=\chi'-i\chi''$
is defined as
  \begin{eqnarray} 
 \chi(\omega) ={1 \over \pi H_0} \int_0^{2\pi}\!\! m(t)\,
   e^{-i \omega t} \, d(\omega t) \,.
  \end{eqnarray}

  In the general nonlinear case the $\chi_\mu (H_0, \omega)$
depend on both $H_0$ and $\omega$. In the useful model
$E(J) = E_c (J/J_c)^n$, the $\chi_\mu(H_0, \omega)$ depend only
on combinations of the form $H_0/\omega^{1/(n-1)}$  or
$\omega/H_0^{n-1}$ or any function of these ratios. Thus,
$\chi(H_0, \omega)$ for different frequencies $\omega$ is
obtained by rescaling the amplitude axis,
$\chi(H_0, c\omega) = \chi(H_0 /c^{1/(n-1)}, \omega)$ for any
constant $c$.
This scaling to a good approximation applies also to other $E(J)$
laws if these are sufficiently nonlinear and if the effective
exponent $n$ is defined as $n=\partial (\ln E)/\partial (\ln J)$
taken at $J=J_c$ where $J_c$ is the typical current density
of the experiment. The nonlinear susceptibility thus depends only
on {\it one} variable combining amplitude and frequency, and further
on an effective exponent $n$ and on the geometry.
   \\[.5cm]
{\bf \large 3. Some Linear AC Susceptibilities}

   The linear susceptibilities $\chi(\omega) = \chi' -i\chi''$
or permeabilities $\mu(\omega) = \chi(\omega) +1$ of conductors
with arbitrary shape and arbitrary complex ac resistivity
$\rho_{ac}$ or penetration depth $\lambda_{ac} =
[ \rho_{ac}(\omega) / i\omega\mu_0 ]^{1/2}$ in a homogeneous
magnetic field $H_a(t) = H_0 e^{i\omega t}$ may be obtained by
solving a linear diffusion equation for $J({\bf r},t)$ or
$H({\bf r},t)$  with diffusivity $D= \rho_{ac} /\mu_0$. In this
way one finds for infinite slabs (width $2a$) in a parallel
field (Kes {\it et al.} 1989) and long cylinders
(Clem {\it et al.} 1976) or spheres of radius $a$ (London 1961)
in an axial field:
  \begin{eqnarray} 
  \chi_{\rm slab}(\omega) = {\tanh u \over u} -1, \\
  \chi_{\rm cyl}(\omega)= {2 {\rm I}_1(u) \over
                             u {\rm I}_0(u) } -1, \\
  \chi_{\rm sphere}(\omega)= {3 \coth u \over u} -{3\over u^2} -1,
  \end{eqnarray}
where $u = a /\lambda_{ac} = [i\omega \mu_0 a^2/
 \rho(\omega) ]^{1/2}$ is complex and the definition
$\chi(\omega) = m(\omega) /| m(\omega \to \infty) |$ was used.
$I_0$ and $I_1$ are modified Bessel functions. At high frequencies
($|u| \gg 1$) one has for $\mu=\chi+1$:  $\mu_{\rm slab} \approx 1/u$,
$\mu_{\rm cyl} \approx 2/u$, and $\mu_{\rm sphere} \approx 3/u$.

  Interestingly, long cylinders in perpendicular $H_a(t)$ yield
the same normalized susceptibility $\chi_{\rm cyl}(\omega)$ (5)
as cylinders in parallel field, but the magnetic moment is twice
as large: $m_{\rm cyl}^\perp (t) = 2 m_{\rm cyl}^\| =
 2\pi a^2 H_a(t) \chi_{\rm cyl}(\omega)$. Since this response
is linear, one may superimpose both solutions and thus finds the
magnetic moment for long cylinders in an applied field $H_a(t)$
inclined at an arbitrary angle $\theta$ away from the cylinder axis,
${\bf m}_{\rm cyl} = ( m_{\rm cyl}^\|,~ m_{\rm cyl}^\perp )$,
  \begin{eqnarray} 
  {\bf m}_{\rm cyl}(t; \theta) = \pi a^2 H_a(t)\,
    \chi_{\rm cyl}(\omega) ( \cos\theta,~ 2\sin\theta ) .
  \end{eqnarray}
The current density in the cylinder is
${\bf J} = (J_\varphi,~J_\|)$,
  \begin{eqnarray} 
  {\bf J}(r,\varphi,t; \theta) =
   {H_a(t) \over \lambda_{ac}} {I_1(r/\lambda_{ac}) \over
   I_0(a/\lambda_{ac}) } ( \cos\theta,~ 2\sin\theta \cos\varphi) .
  \end{eqnarray}

 In general, the linear susceptibility for any geometry may be
written as an infinite sum, or approximated by a finite sum,
of the form
  \begin{equation}    
  \chi(\omega) = -w \sum_\nu {\Lambda_\nu b_\nu^2 \over
   w +\Lambda_\nu} \Big/ \sum_\nu \Lambda_\nu b_\nu^2 \,.
  \end{equation}
Here $\Lambda_\nu$ ($\nu=1$, 2, $\dots$) are the eigenvalues
of an eigenvalue problem, the $b_\nu$ (``dipole moments'') are
integrals over the eigenfunctions $f_\nu({\bf r})$, and the complex
variable $w$ is proportional to $i\omega /\rho_{ac}(\omega)$.
 The sum in the denominator of Eq.\ (9) provides the
normalization $\chi(\omega \to \infty) = -1$ (ideal diamagnetic
screening).

  For thin film disks and strips with diameter or width $2a$ and
thickness $2b \ll 2a$ in perpendicular field the $\Lambda_\nu$
and $b_\nu$ are tabulated by K\"otzler {\it et al.} 1994 and
Brandt 1994c, and the main variable is
  \begin{eqnarray} 
    w = {i\omega \mu_0 ab \over \pi \rho_{ac}(\omega)}
      = {ab \over \pi \lambda_{ac}^2}
      = i \omega \tau(\omega)\,.
  \end{eqnarray}
Tables of the $\Lambda_\nu$ and $b_\nu$ for thick disks or
cylinders of arbitrary length in axial field are given by
Brandt 1998. For infinite bars with rectangular cross-section
$|x| \le a$, $|y| \le b$, in perpendicular $H_a(t)$ along $y$,
the variable $w$ (10) applies also and the eigenvalue problem reads
(Brandt 1996b)
  \begin{eqnarray} 
  f_\nu ({\bf r}) = - {\Lambda_\nu \over ab}
  \int_{-a}^a \!\!\! dx' \int_{-b}^b \!\!\! dy' \ln|{\bf r-r'}|
                \, f_\nu ({\bf r'}) ,
  \end{eqnarray}
where ${\bf r} = (x,y)$, ${\bf r'} = (x',y')$. With the
normalization $\int f_\mu f_\nu d^2r = \delta_{\mu \nu}$ the
``oscillator strengths'' are in this case
  \begin{eqnarray} 
  b_\nu = {\pi \over ab} \int \! x f_\nu(x,y) \,d^2 r \,.
  \end{eqnarray}
For practical purposes, a finite number of terms $\nu=1 \dots N$
in the sum (9) is sufficient. When the (real and positive) numbers
$\Lambda_\nu$ and $b_\nu$ are known for a given geometry, then
$\chi(\omega)$ (9) may be calcuated for any complex resistivity
$\rho_{ac}(\omega)$. By inverting this relationship between the
two complex functions $\chi(\omega)$ and $\rho_{ac}(\omega)$
numerically, the complex resistivity $\rho_{ac}(\omega)$ may be
obtained from measured ac susceptibilities as done by
K\"otzler {\it et al.} 1994.

   The linear $\chi'(\omega)$ and $\chi''(\omega)$ of diks or
cylinders with Ohmic resistivity $\rho_{ac}(\omega)=\rho=$const
in axial $H_a(t)$ are shown in Fig.\ 1 for various aspect ratios
$b/a=0 \dots 10$ ($a$ = radius, $2b$ = length) versus the reduced
frequency $\omega\tau$ with $\tau=\mu_0 a b /(\pi\rho)$.
Note that in this double logarithmic plot for sufficiently thin
disks ($b\ll a$) the $\mu(\omega)=\chi(\omega)+1$ with increasing
frequency crosses over from the behavior
in perpendicular geometry ({\it nonlocal} flux diffusion)
(Brandt 1994c),
  \begin{eqnarray}    
   \mu' = \chi'+1 \propto 1/\omega,~~
   \mu''= \chi''\propto \ln({\rm const}\cdot\omega\tau)/\omega\,,
  \end{eqnarray}
to the parallel geometry ({\it local} flux diffusion),
  \begin{eqnarray}    
  \mu' \approx \mu''\propto 1/\sqrt \omega \,.
  \end{eqnarray}
At all aspect ratios $b/a>0$
the real and imaginary parts at large frequencies coincide,
$\mu'(\omega)=\mu''(\omega)$. The physical
reason for this finding is that above the frequency where the skin
depth $\delta=\sqrt2 \lambda_{ac} = (2\rho/i\omega \mu_0)^{1/2}$
coincides with the half thickness $b$, the Ohmic conductor nearly
behaves like an ideal diamagnet (or superconductor in the Meissner
state), screening almost all magnetic flux from the interior of
the conductor. The magnetic field lines thus have to flow around
the bar or cylinder such that ${\bf B}$ is {\it parallel} to the
specimen surface everywhere. Therefore, at high frequencies, thin
(and thick) Ohmic (and non-Ohmic) conductors in a perpendicular
magnetic ac field {\it behave as if the field were applied parallel}.
   \\[0.5cm]
{\bf \large 4. Nonlinear Magnetic Response of Thin Films}

  The nonlinear susceptibility of thin superconductor films with
given material law in a perpendicular magnetic field, in general may
be obtained only numerically, e.g.\ by time-integrating an integral
equation for the current
density $J({\bf r},t)$, inserting this into Eq.\ (1) for the
magnetic moment $m(t)$, and then evaluating the Fourier
integral (2) for $\chi(H_0,\omega)$. Such results are presented
for thin film disks and rings by Brandt 1997b and for
thick disks (or short cylinders) by Brandt 1998.

  For narrow rings or thin-walled hollow cylinders, analytical
expressions for $\chi(H_0)$ are available in the Bean limit
(Brandt 1997b). In this case (and also when a ring contains one or
several weak links) the virgin magnetization curve is a straight
line, $m(H_a) = -H_a m_{\rm sat}/H_p$ (ideal screening or Meissner
state) which saturates to a horizontal line, $|m| = m_{\rm sat}$
at $|H_a|  \ge H_p\,$, the field of first penetration of flux
into the hole. This is so since the current in the ring is limited
to a critical value $I_c$.
For a narrow ring with width $w=a-a_1 \ll R =(a+a_1)/2$
($a$, $a_1$ are the outer and inner radius) and thickness
$d \ll R$ one has $I_c = J_c w d$, $m_{\rm sat}=\pi R^2 I_c$,
$H_p \approx I_c \ln(5R/x)/\pi R$ where the inductivity
$L \approx \mu_0 R \ln(5R/w)$ was used (Brandt 1997b). For such rings
or tubes, in cycled $H_a(t) = H_0 \sin\omega t$ the magnetization
loop at small amplitudes $H_0 \le H_p$ degenerates to a straight line,
yielding a constant and real $\chi(H_0) = - m_{\rm sat} / H_p\,$,
while at large  $H_0 \ge H_p$  the loop $m(H_a)$ is a parallelogramme
with height $2m_{\rm sat}$. This yields the normalized ac
susceptibility $\chi(h) = \chi(H) H_p/m_{\rm sat} = \chi'-i\chi''$,
with $h = H_0/H_p$ and $s = 2/h\, -1$,
  \begin{eqnarray} 
  \chi'(h) =& -1,~~~                           &~ \nonumber\\
  \chi''(h)=&  0,                  &~~~h \le 1\,, \nonumber\\
  \chi'(h) =& -{1\over2} -{1\over\pi} \arcsin s - {1\over\pi}
              s \,\sqrt{1 -s^2}\,,  &~            \nonumber\\
  \chi''(h)=& {\displaystyle {4\over\pi} {h-1 \over h^2}
           =  { 1-s^2\over \pi } },  &~~~h \ge 1\,,
  \end{eqnarray}
see the curve $a_1/a =1$ in Fig.\ 2.
The polar plot ($\chi''$ versus $\chi'$ with $h$ as parameter) of the
ring susceptibility (15) is {\it symmetric}, i.e.\
$\chi''(\chi')$ yields the same curve as $\chi''(-1-\chi')$.
The maximum of $\chi''_{\rm max} = 1/\pi = 0.318$ occurs at
$h=2$ (at $s=0$). For large amplitudes $h=H_0/H_p \gg 1$ one has
$\chi'(h) \approx -1.69 /h^{3/2}$ and $\chi''(h) \approx 4/(\pi h)$.

   Very good approximate expressions are available within the
Bean critical state model for thin strips (width $2a$, thickness
$d \ll a$, length $L \gg a$), disks (radius $a$), and ellipses
(semi-axes $a$ and $b$, excentricity $e=b/a \le 1$)
(Mikitik and Brandt 1999).
In these three cases and also for square and rectangular films,
the virgin magnetization curve with less than $1\%$ error is
  \begin{eqnarray}    
  m(H_a) =  -m_{\rm sat}\, \tanh (H_a / H_1) \,,
  \end{eqnarray}
with $H_1 = m_{\rm sat} /|m'(0)|$, and explicitly,
  \begin{eqnarray}    
  m_{\rm sat} = J_c d L a^2, &~~H_1 =J_c d/\pi& ~~~~~{\rm (strip)},\\
  m_{\rm sat} = {\pi\over 3}J_c da^3, &~~H_1 = \pi J_c d/8&
                                             ~~~~~~{\rm (disk)},\\
  m_{\rm sat} = {4\over 3} J_c da b^2 { \cos(e\pi/2) \over 1-e^2 },
  &~~H_1 = J_c d { E(k) \over \pi} { \cos(e\pi/2) \over 1-e^2 }&
                                         ~~~{\rm (ellipse)}.
  \end{eqnarray}

 The formulae (19) for ellipses with $e=b/a \le 1$ reproduce the
results for disks ($e \to 1$, radius $a=b$) and strip ($e\to 0$,
length $2a$, width $\sqrt{8/3} \,b$). From the virgin curve $m(H_a)$
 (16), the Bean magnetization loops with amplitude $H_0$ follow as
  \begin{eqnarray}    
   m_\downarrow(H_a, H_0) = m(H_0) + 2m\Big( {H_a-H_0 \over 2} \Big)
  \end{eqnarray}
(decreasing $H_a$), and
$m_\uparrow(H_a, H_0) = -m_\downarrow(-H_a, H_0)$ (increasing $H_a$).
Inserting this into definition (2) one obtains the universal
nonlinear ac susceptibility $\tilde \chi(h)$ of films normalized
to $\tilde \chi(0) = -1$ and depending on $h = H_0 /H_1$:
  \begin{eqnarray} 
  \tilde\chi(h) ={2i \over \pi h} \int_{-\pi/2}^{\pi/2} \!
  \Big[ 2\tanh \Big( h{1-\sin\varphi \over 2} \Big)
    - \tanh\, h \Big] \, e^{-i \varphi} \,d\varphi \,.
  \end{eqnarray}
From this universal function the not normalized $\chi(H_0)$ is
obtained as
  \begin{eqnarray} 
   \chi(H_0) = { m_{\rm sat} \over H_1 }\,
   \tilde\chi\Big( {H_0 \over H_1} \Big)\,.
  \end{eqnarray}
Figure 2 shows some normalized ac susceptibilities $\tilde \chi(h)$
for thin film rings and for the disk and strip in the Bean limit,
i.e.\ for large creep exponent $n \to \infty$. Note that
the curves for disks and strips practically coincide.

  The nonlinear ac susceptibility of disks with various creep
exponents $n \ge 3$ is depicted in Fig.\ 3. Here
$E(J)=E_c (J/J_c)^n$ was assumed, the circular frequency was
$\omega = \omega_1 \equiv 2E_c /(\mu_0 J_c da)$, and the amplitude
$H_0$ is in units $J_c d$ ($d$ = tickness, $a$ = radius of the disk).
The plots in Fig.\ 3  apply to any frequency
$\omega$ if the  unit of $H_0$ is changed to
$ (\omega/\omega_1)^{1/(n-1)} J_c d$.
   \\[0.5cm]
{\bf \large 5. Final Remarks}

  The measured ac susceptibilities depend on the temperature $T$
{\it indirectly} via the $T$ dependence of the material properties.
The {\it nonlinear} $\chi(H_0, \omega)$ depends on $J_c(B,T)$
and on the creep exponent $n(B,T) = U_0 /kT$, see above.
The {\it linear} $\chi(\omega)$ depends on the variable $w$,
Eq.\ (10), which contains the linear complex ac resistivity
$\rho_{ac} (\omega, B, T)$, see e.g.\ Coffey and Clem 1990 and
Brandt 1990 for theories of $\rho_{ac}(\omega)$ of
type-II superconductors. The linear $\chi(\omega)$ usually is
measured in a bias dc magnetic field $H_a^{dc} \gg H_0$, but the
nonlinear $\chi(H_0,\omega)$ discussed above assumed the absence of
a bias field and has to be recomputed if $H_a^{dc} >0$.

  Furthermore, if $H_a(t)$ is too small, the effects of a finite
$H_{c1}$ should be considered, which were disregarded here but may
lead to a geometric barrier, in particular in thin films, as discussed
in the introduction. The edge barrier may be suppressed by using
small measuring coils close to the film, see e.g.\
Gilchrist and Brandt 1996 for a detailed theory and analytic formulae.
The numerical program described by Brandt 1999 in principle accounts
for arbitrary reversible magnetization $H(B)$ and thus for $H_{c1}$
and for the edge barrier in strips or disks of constant thickness.
It may also be generalized to allow for dc and ac transport currents
in strips.

  The problem of ac losses in thin strip superconductors with
transport current is related to the problem of ac susceptibilities.
It is conceptually difficult since with increasing ac amplitude
the losses may be caused first by flux jumping over the edge
barrier, then by bulk pinning (Bean critical state), then (after
full penetration) by the rapidly increasing electric field
$E \propto J^n$. The hysteresis losses caused by the edge barrier in
superconductor strips without bulk pinning were recently calcuated
by Clem and Benkraouda 1998.
   \\[0.9cm]
{\bf \large 5. References}

\begin{description}  \itemsep -2.pt \parsep 0pt

\item Bean, C.~P., 1964, {\it Rev.\ Mod.\ Phys.}, {\bf 36}, 31.
\item Bean, C.~P., and Livingston, J.~D., 1964,
  {\it Phys.\ Rev.\ Lett.}, {\bf 12}, 14.
\item Blatter G., Feigel'man M.~V, Geshkenbein, V.~B., Larkin, A.~I.,
   and Vinokur, V.~M., 1994, {\it Rev.\ Mod.\ Phys.}, {\bf 66} 1125.

\item Brandt, E.~H., 1991, {\it Phys.\ Rev.\ Lett.}, {\bf 66}, 3213. 
\item Brandt, E.~H., 1994a, {\it Phys.\ Rev.\ B}, {\bf 49}, 9024. 
\item Brandt, E.~H., 1994b, {\it Phys.\ Rev.\ B}, {\bf 50}, 4034. 
\item Brandt, E.~H., 1994c, {\it Phys.\ Rev.\ B}, {\bf 50}, 13833. 
\item Brandt, E.~H., 1995a, {\it Phys.\ Rev.\ B}, {\bf 52}, 15442. 
\item Brandt, E.~H., 1995b, {\it Rep.\ Prog.\ Phys.}, {\bf 58}, 1465.
\item Brandt, E.~H., 1996a, {\it Phys.\ Rev.\ Lett.}, {\bf 76}, 4030. 
\item Brandt, E.~H., 1996b, {\it Phys.\ Rev.\ B}, {\bf 54}, 4246. 
\item Brandt, E.~H., 1997a, {\it Phys.\ Rev.\ Lett.}, {\bf 78}, 2208. 
\item Brandt, E.~H., 1997b, {\it Phys.\ Rev.\ B}, {\bf 55}, 14513. 
\item Brandt, E.~H., 1998, {\it Phys.\ Rev.\ B}, {\bf 58}, 6506, 6523.
\item Brandt, E.~H., 1999, {\it Phys.\ Rev.\ B}, {\bf 59}, 3369

\item Brandt, E.~H., and Gurevich, A., 1997, {\it Phys.\ Rev.\ Lett.},
  {\bf 76}, 1723.
\item Brandt, E.~H., and Indenbom, M., 1993,
      {\it Phys.\ Rev.\ B}, {\bf 48}, 12893.
\item Brandt, E.~H., Indenbom, M., and Forkl, A., 1993,
      {\it Europhys.\ Lett.}, {\bf 22}, 735.
\item Campbell, A.~M., and Evetts, J.~E., 1972, {\it Adv.\ Phys.},
  {\bf 72}, 199.
\item Clem, J.~R., and Benkraouda, M., 1998, {\it Phys.\ Rev.\ B},
   {\bf 58}, 15103.
\item Clem, J.~R., Kerchner, H.~R., and Sekula, T.~S., 1976,
   {\it Phys.\ Rev.\ B}, {\bf 14}, 1893.
\item Coffey, M., and Clem, J.~R., 1991,
   {\it Phys.\ Rev.\ Lett.}, {\bf 67}, 386.

\item M.~J.~W.~Dodgson, M.~J.~W., Geshkenbein, V.~B., Nordborg, H.,
  and G.~Blatter, 1998, {\it Phys.\ Rev.\ Lett.}, {\bf 80}, 837;
  {\it Phys.\ Rev.\ B}, {\bf 57}, 14498.
\item Doyle, T.~B., Labusch, R., and Doyle, R.~A., 1997,
             {\it Physica C}, {\bf 290}, 148.

\item Gilchrist, J., and Brandt, E.~H., 1996, {\it Phys.\ Rev.\ B},
   {\bf 54}, 3530.
\item Gilchrist, J., and Dombre, T., 1994, {\it Phys.\ Rev.\ B},
   {\bf 49}, 1466.
\item Gurevich, A., and Brandt, E.~H., 1994, {\it Phys.\ Rev.\ Lett.},
   {\bf 73}, 178.
\item Gurevich, A., and Brandt, E.~H., 1997, {\it Phys.\ Rev.\ B},
   {\bf 55}, 12706.

\item Kes, P.~H., Aarts, J., van den Berg, J., van der Beek, C.~J.,
   Mydosh, J.~A., 1989, {\it Supercond.\ Sci.\ Technol.}, {\bf 1}, 242.
\item K\"otzler, J., Nakielski, G., Baumann, M., Behr, R., Goerke, F.,
   and Brandt, E.~H., 1994, {\it Phys.\ Rev.\ B}, {\bf 50}, 3384.
\item Labusch, R., and Doyle, T.~B., 1997, {\it Physica C}, {\bf 290},
   143.
\item London, F., 1961, {\it Superfluids II. Macroscopic Theory of
   Superconductivity} (New York: Dover). p. 35.

\item Majer, D., Zeldov, E., and Konczykowski, M., 1995,
  {\it Phys.\ Rev.\ Lett.}, {\bf 75}, 1166.
\item McElfresh, M., Zeldov, E., Clem, J.~R., Darwin, M., Deak, J.,
    and Hou, L., 1995, {\it Phys.\ Rev.\ B}, {\bf 51}, 9111.
\item Mikheenko, P.~N., and Kuzovlev, Yu.~E., 1993, {\it Physica C},
   {\bf 204}, 229.
\item Mikitik, G.~P., and Brandt, E.~H., 1999, {\it Phys.\ Rev.\ B}
   (submitted).

\item Norris, W.~T., 1970, {\it J.\ Phys.\ D: Appl.\ Phys.},
   {\bf 3}, 489.
\item Rhyner, J., 1993, {\it Physica C}, {\bf 212}, 292.
\item Roulin, M., Junod, A., and Walker, E., 1998,
  {\it Phys.\ Rev.\ Lett.}, {\bf 80}, 1722.

\item Sasagawa, T., Kishio, K., Togawa, Y., Shimoyama, J., and
  Kitazawa, K., 1998, {\it Phys.\ Rev.\ Lett.}, {\bf 80}, 4297.
\item Schilling, A., Fisher, R.~A., Phillips, N.~E., Welp, U.,
  Dasgupta, D., Kwok, W.~K, and Crabtree, G.~W., 1996, {\it Nature},
  {\bf 382}, 791.
\item Schuster, Th., Kuhn, H., Brandt, E.~H., and Klaum\"unzer, S.,
   1997, {\it Phys.\ Rev.\ B} {\bf 56}, 3413.

\item  Welp, U., Fendrich, J.~A., Kwok, W.~K., Crabtree, G.~W., and
  Veal, B.~W., 1996, {\it Phys.\ Rev.\ Lett.}, {\bf 76}, 4809.
\item Zeldov, E., Larkin, A.~I., Geshkenbein, V.~B.,
   Konczykowski, M., Majer, D., Khaykovich, B., Vinokur, V.~M.,
   and Shtrikman, H., 1994, {\it Phys.\ Rev.\ Lett.}, {\bf 73}, 1428.
\item Zeldov, E., Majer, D., Konczykowski, M., Geshkenbein, V.~B.,
  Vinokur, V.~M., and Shtrikman, H., 1995, {\it Nature}, {\bf 375}, 373.

\end{description}
\newpage

 \begin{figure}[F1]
\epsfxsize= .75\hsize  \vskip 1.5\baselineskip
\centerline{ \epsffile{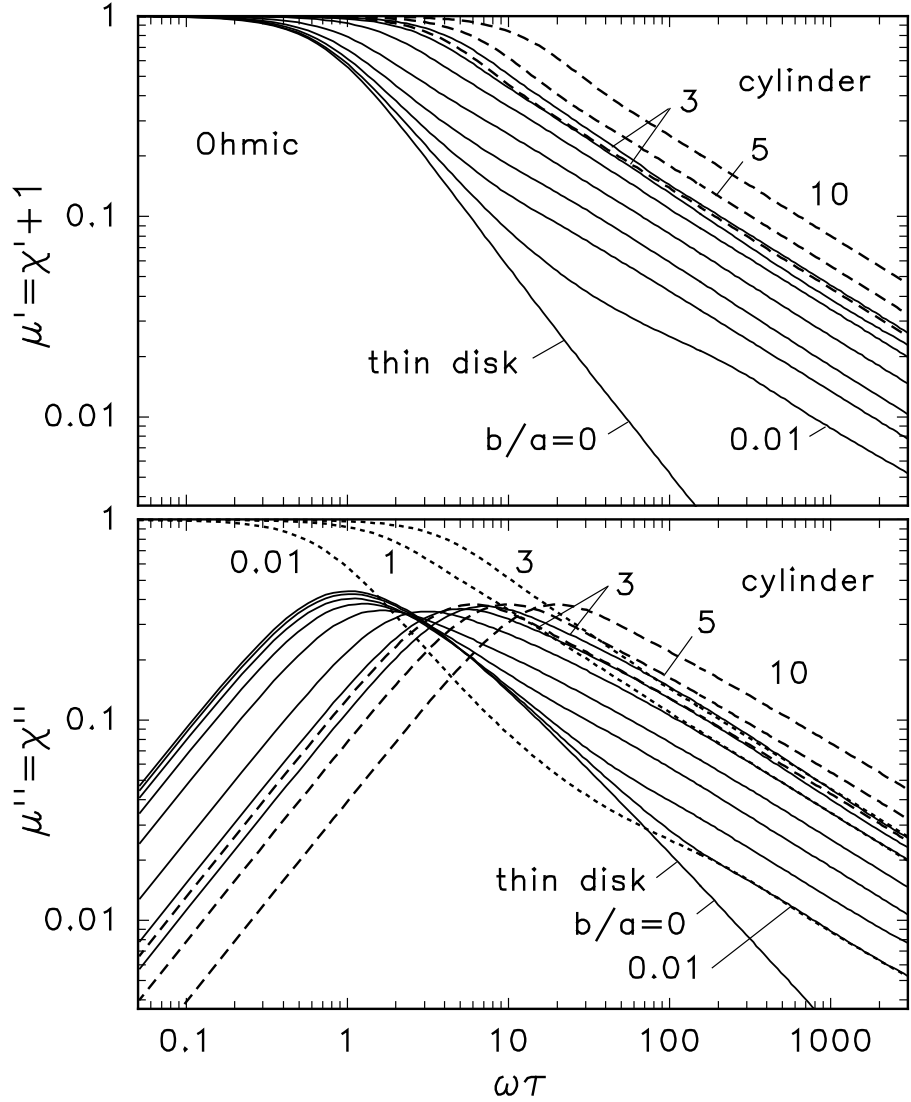}}

\caption{The real (top) and imaginary (bottom) parts of the linear
ac permeability $\mu=\chi+1=\mu' -i\mu''$ of Ohmic cylinders with
rectangular cross-section $2a\times 2b$ in an axial magnetic
ac field for aspect ratios $b/a =0$ (thin disk), $b/a =0.01$, 0.03,
0.1, 0.3, 1, 2, and 3 (short cylinders) computed from
the sum (9) (solid lines).
The dashed curves give the analytic expression (5)
for long cylinders with $b/a =3$, 5, and $10$, with the variable
$u=a/\lambda_{ac} = (i\omega\tau \pi a/b)^{1/2}$ inserted,
where $\tau=\mu_0 a b /(\pi\rho)$ is the time scale.
For $b/a=3$ the numerical and analytical curves are very close.
To facilitate comparison, some $\mu'$ curves are repeated as
dotted lines in the lower plot.
    }
\end{figure}

 \begin{figure}[F2]
\epsfxsize= 0.65\hsize  \vskip 1.5\baselineskip
\centerline{ \epsffile{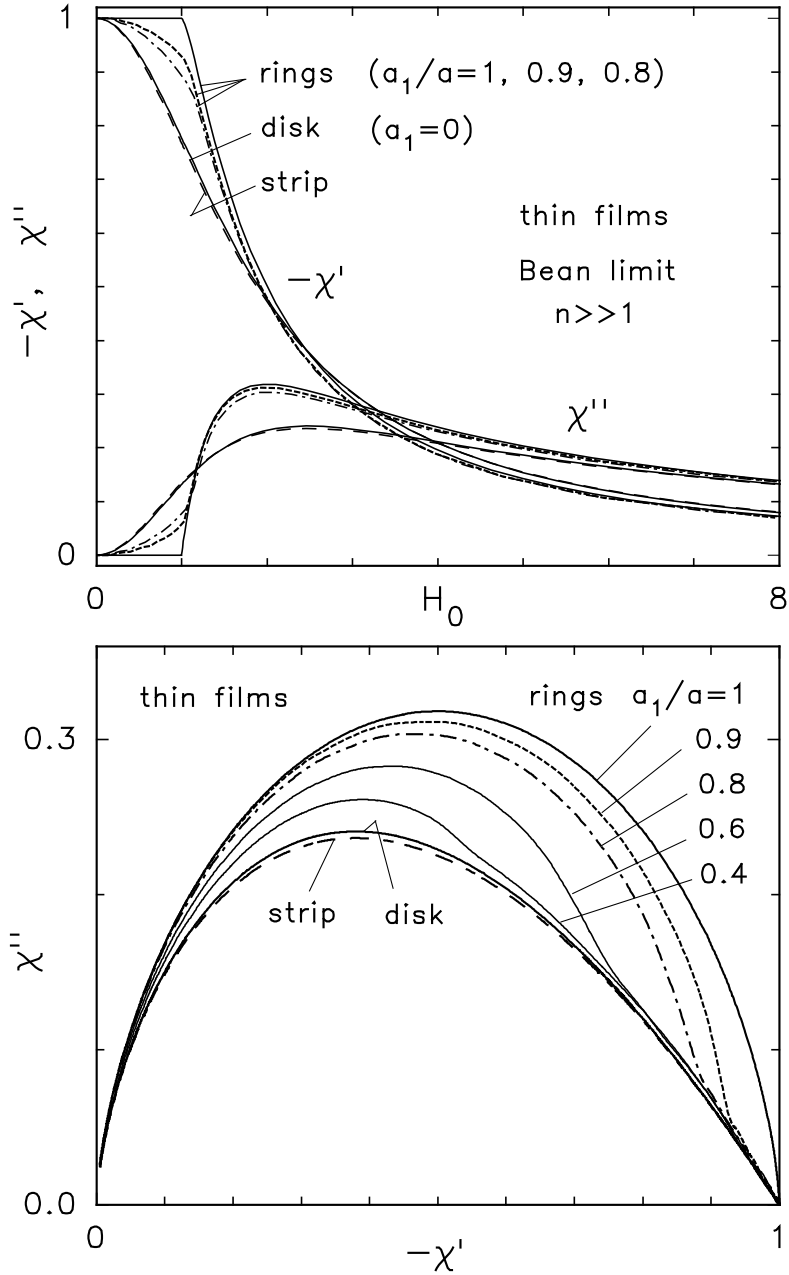}}

\caption[]{ {\it Top:\/} Real and imaginary parts of the
nonlinear complex susceptibility $\chi(H_0) = \chi'- i\chi''$ of
thin strips, disks, and rings with various ratios $a_1/a$ of the
inner and outer radius, plotted versus the
amplitude $H_0$ of the ac field. Bean model with constant $J_c$.
The unit of $H_0$ is $H_1= J_c d/\pi$ for the strip
and $H_1=\pi J_c d /8$ for the disk. For rings $H_0$ is in units
of a penetration field $H_p' \approx H_p$ which gives best fit
to the sharp rise in $\chi''$ of the ideal ring:
$H_p'/J_c d = 0.126$ (0.210) for $a_1/a = 0.9$ (0.8), while from
Table 1 in Brandt 1997b  one has $H_p /J_c d = 0.136$ (0.239).
For the narrow ring with width $w=a-a_1 \ll a$ one has
$H_p' = H_p$, $H_p/J_c d \approx (w/ \pi a) \ln(5a/w)$.
  {\it Bottom:\/}  Polar plots of the same $\chi = \chi'-i\chi''$.
For the narrow ring $\chi' $ and $\chi''$ [Eq.\ (15)] are
discontinuous at $H_0 = H_p$ and the polar plot is symmetric.
    }
\end{figure}

 \begin{figure}[F3]
\epsfxsize= 0.65\hsize  \vskip 1.5\baselineskip
\centerline{ \epsffile{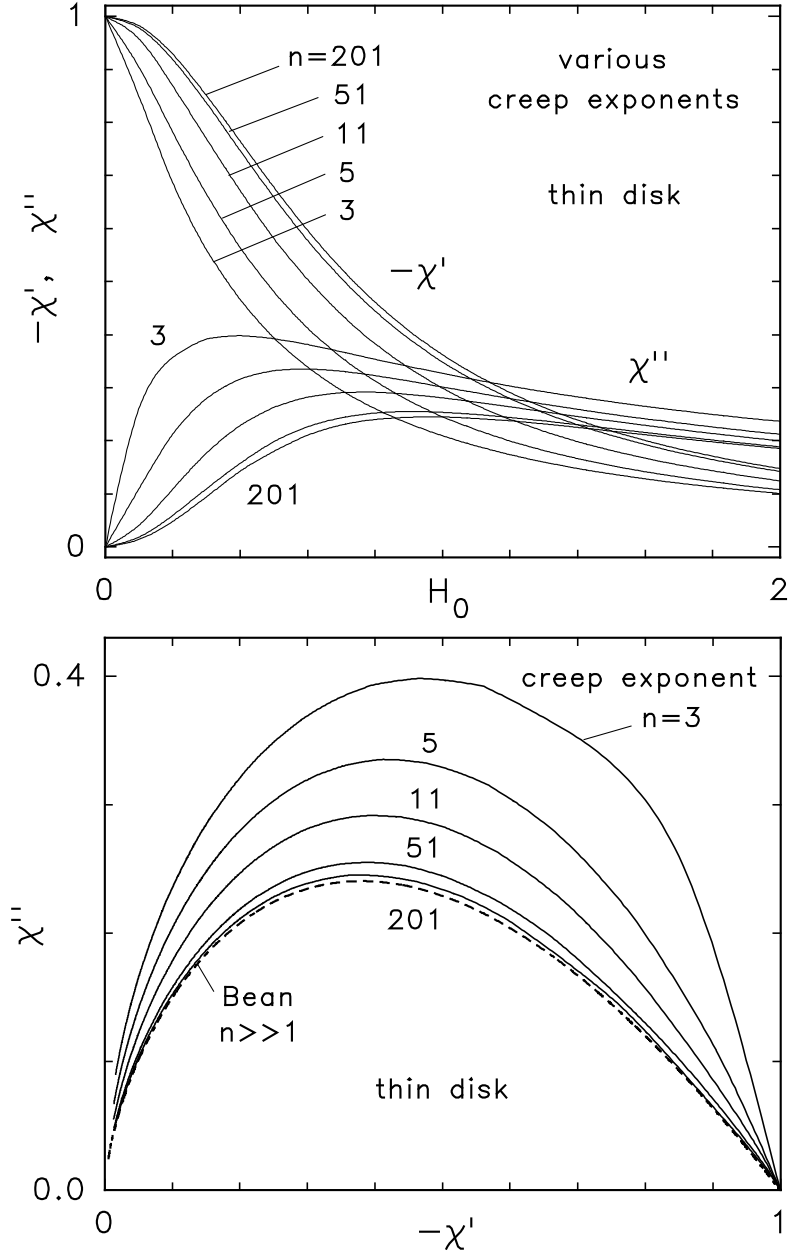}}

\caption[]{{\it Top:\/} Real and imaginary parts of the
nonlinear complex susceptibility $\chi(H_0) = \chi'- i\chi''$ of a
thin disk plotted versus the ac amplitude $H_0$ for various creep
exponents $n=3$, 5, 11, 51, and 201 in the law $E=E_c (J/J_c)^n$ with
constant $J_c$. Here $H_0$ is in units $J_c d$.
{\it Bottom:\/}  Polar plots of the same $\chi = \chi'-i\chi''$.
The dashed line shows the Bean model ($n \to \infty$) for the disk.
  }
\end{figure}
\end{document}